# DESIGN AND IMPLEMENTATION FOR AUTOMATED NETWORK TROUBLESHOOTING USING DATA MINING

Eleni Rozaki

School of Computer Science & Informatics, Cardiff University, UK

## ABSTRACT

*The efficient and effective monitoring of mobile networks is vital given the number of users who rely on such networks and the importance of those networks. The purpose of this paper is to present a monitoring scheme for mobile networks based on the use of rules and decision tree data mining classifiers to upgrade fault detection and handling. The goal is to have optimisation rules that improve anomaly detection. In addition, a monitoring scheme that relies on Bayesian classifiers was also implemented for the purpose of fault isolation and localisation. The data mining techniques described in this paper are intended to allow a system to be trained to actually learn network fault rules. The results of the tests that were conducted allowed for the conclusion that the rules were highly effective to improve network troubleshooting.*

## 1. INTRODUCTION

Network management is a difficult task regardless of the equipment that is part of the network. However, the difficulties of network management are increased when devices and equipment from different manufacturers are used. With the use of a variety of devices on a network, the process of anomaly detection problems and issues becomes more difficult because the circuits in the network do not have a single console under which they operate [21]. What would be helpful would be a monitoring process that utilises a single console.

In this paper, an investigation is undertaken of a proposed method automated scheme for network troubleshooting and fault detection based on data mining techniques to provide converged services with functionalities that facilitate its fault-handling and operational management. With the use of Weka classifiers (rules, trees, Bayes), one of the main challenges is to carry out effective and accurate monitoring of services with the detection and classification of possible network anomalies that may arise at runtime so that appropriate measures rules and decision trees can be made to ensure correct fault-handling and compliance with the Quality of Service (QoS) constraints established between service providers and end-users [1].

The current process that is used to detect and diagnose faults in communications networks is manual process that requires a technician to address problems once they have created inefficiencies for users. Unfortunately, automated methods for detecting and diagnosing network faults do not exist, or are, at best, very early in the development process [21]. The fact that automated fault detection methods do not exist is an important gap within both the practical and academic literature. The goal of this investigation is to provide an efficient and effective means by which the process of detecting and diagnosing network anomalies can be automated.

The FCAPS framework was defined and implemented in 1997 as a management framework for network provides. The FCAPS framework stands for fault, configuration, accounting, performance, and security. Within this framework, performance management is the process of quantifying, measuring, analysing, and controlling the performance of the components of a network. Fault management is the process of acquiring, revealing, and then counteracting faults that exist in the network. Configuration management is the process of obtaining and controlling the configuration parameters of a network. Accounting management refers to usage statistics, cost allocation, and pricing. Security management is the process of monitoring access to a network based on defined policies [2].

We are working in the diagnosis stage and investigating how to automate the process of setting up a fault classification algorithm based on KPI alarms to identify network anomalies. At the same time, we investigate classification rules and decision tree algorithms between different alarms of KPIs based on causes and symptoms of network malfunctions and demonstrate both system performance and localisation of the KPI alarms. [3].

Within the academic literature, a great deal of time and effort has been devoted to methods of fault management and anomaly diagnosis within wireless networks. A variety of methods, such as hierarchical support vector machine, the creation of assessment indexes, and neural algorithms based on faculty connectors [3][4][5].

There has been a limited amount of research regarding the degree of fault alarm under data mining techniques. This is problematic given that defect management on telecommunication networks has been studied extensively utilizing the correlation techniques. In this way, the current paper helps to move forward the existing knowledge regarding anomaly detection and diagnosis.

## 2. TROUBLESHOOTING MANAGEMENT USING DATA MINING

Network administrators need to recognise at the earliest possible moment that a network communication failure is occurring. However, with current methods, network operators must conduct manual searchers through large amounts of data in order to find the problems. Furthermore, as mobile networks are often handled through different facilities, each facility is likely to have different traffic characteristics and anomalies, as well as different faults. The entire process is not only complicated, but can require hours of work in order to return a network to full optimisation. What is needed is a process by which fault detection rules can be extracted automatically based on historical fault data. The creation of an automated means of fault detection may be accomplished with the use of decision trees that can allow for rules to be generated based on the issues that are present at a given time. The methodology that is used is considered an inference engine for network troubleshooting as shown in Figure 1. The engine maintains a list of metric values and the potential causes for the fault history. Based on the data and the interpretation of the user, the engine can perform calculations on the correlation between the metrics and the faults that are occurring. Then, detection rules can be applied [21].

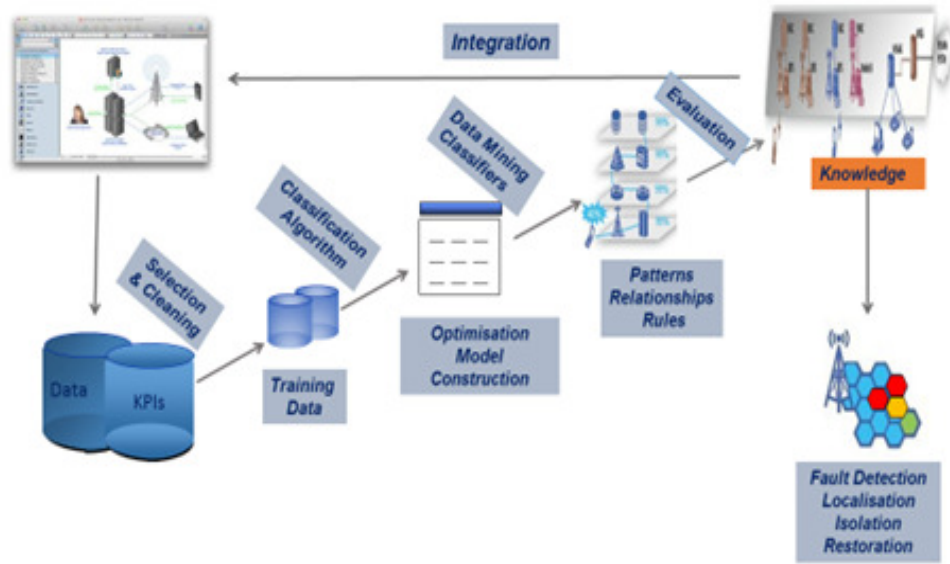

Figure 1.  Troubleshooting management using data mining

The process of mining data so that it can be gathered and used to make decisions does not involve a single activity. Instead, data mining involves a series of steps that must be undertaken in order to allow large amounts of data to be reduced and put into a format that can be used for the purpose of decision-making [7].

The first step in the data mining process is known as data cleaning. In this step, data are examined to determine which data are relevant for the given task or purpose. Irrelevant data and data that may be considered to be invalid are removed. Next data integration occurs in which data from multiple sources are combined into a single source or dataset. It is often the case that the data that are combined are heterogeneous [8].

The next step in the data mining process is data selection in which further examination of the data occurs in order to determine which data should be taken from the dataset to be used for the task at hand. Once the data selection has been completed, then data transformation and training can occur. This stage in the data mining process involves transforming the selected data into appropriate formats for the mining procedure. [8].

The following step in the process is the actual data mining. In this step, various techniques for analysing data are used. Then, pattern evaluation is undertaken in which interesting or unique patterns in the data are identified. The identification of patterns occurs in relation to the larger issue or problem that is attempting to be addressed [8].

With the use of Weka, the final step in the data mining process is knowledge representation. This process of incorporating Weka tools into the data process requires analysing and pre-processing the data, as well as examining the validity of the key performance indicators. Moreover, class attributes must be defined followed by the actual extraction of the features that will be used for classification. Consequently, a subset of features is selected to be used as part of the knowledge construction [8].

The following part of the knowledge representation using Weka is the investigation of any imbalances in the selected data and a determination of how those imbalances may be counteracted. Additionally, a subset of instances is chosen. The subset of instances might be the records upon which the learning will be based. Following the selection of the subset of instances, the classifier algorithm is applied for the learning process. Finally, a testing method is chosen in order to estimate the performance of the algorithm [8].

## 2.1 Classifier Training Methods

The ability to distinguish between different classes of faults can only occur if data for all of the different issues that can occur must be available. The fault detection methodology that is used is only able to function based on the data that are provided. The system must learn the anomalies that have occurred in the past in order to be able to detect different types of issues in the present. By providing data for different types of faults, the neural net classifier can learn differences between normal activities in a network and activities that are not normal. With more historical fault data, the system can actually achieve higher rates of effectiveness at detecting real problems, as well as reducing the rate of false alarms [22].

## 3. DECISION TREE AND DECISION RULES ALGORITHMS

A decision is defined as a structure that is used to classify data that have common attributes [21]. A decision tree represents a single rule around which data are categorised. Within a decision tree, there are internal nodes, leaf nodes, and edges. Internal nodes are specific attributes around which data are partitioned. Leaf nodes are labels for categorization of the data. Edges are labels for possible conditions of the attributes contained in the parent node. Alternatively, a decision can be made regarding a set of rules generated for selected variables. Research has shown that decision trees and decision rules are highly effective for unbalanced classification problems, which means that the abnormal class has only a few training examples [21].

There are a large number of tools that can be used in data mining to handle a variety of tasks, such as data pre-processing, regression, clustering, feature selection, and even visualization. However, it is important to understand that the various tasks that can be performed with data mining tools require different algorithms. At the same time, a programming environment is needed to run the algorithms [9]. In this study, the Waikato Environment for Knowledge analysis (Weka) has been chosen because of its ability to run a variety of different algorithms and its general robustness [10].

An important part of this study is understanding the classifiers that are used for the network anomaly detection. The classifiers that are of importance to this study are briefly explained as follows:

### 3.1. J48 tree

The J48 tree functions through the creation of decision trees in which data are examined with the goal of reducing the data into small subsets under the smallest subset can be found at which point an optimised outcome has been determined. A leaf node is created within the decision tree to indicate that a particular class should be chosen [9][11].

### 3.2. LAD Tree

The Logical Analysis of Data (LAD) tree works by creating a classifier for a binary target. The classifier is created through a process in which a logical expression is learned that has the ability to determine positive and negative samples in a dataset. The underlying theory of the LAD tree is that any binary point that contain some positive patterns but not negative patterns could be classified as positive. At the same time, any binary point that has negative patterns but no positive patterns can be classified as negative. In a similar manner to the J48 tree, a large set of patterns occurs and then subsets are determined until a subset is determined that satisfies the patterns in the data [11].

### 3.3. JRip

The JRip classifier, which was introduced in 1995 by W. W. Cohen, is an optimised version of an older classifier that Cohen created [12]. JRip was implemented with the prepositional rule learner known as Repeated Incremental Pruning to Produce Error Reduction (RIPPER), meaning that it has the ability to replace or revise rules. With JRip, it is possible to isolate some of the data being examined for training. At the same time, a decision can be made regarding the set of rules generated for selected attributes.

### 3.4. PART

The PART algorithm [24] is an algorithm that does not function to generate accurate rules through global optimisation. Instead, PART functions to build a rule and then remove the instances that it covers in order to create a recursive rule until there are no instances that remain. Another way of thinking about the PART algorithm is that creates decision lists. With the decision lists, new data are compared to the rules in the decision list. Upon the first match of a rule, data are assigned to that rule. In this way, PART makes the best leaf into a rule rather than the optimised leaf [9].

## 4. BAYESIAN NETWORKS

Many researchers have investigated the use of Bayesian networks for fault detection, either with other network troubleshooting techniques or as a means of improving other anomaly detection methods. However, the research emphasis on the use of Bayesian networks for troubleshooting management has primarily been on anomaly diagnosis as opposed to actual fault detection. Bayesian networks represent causal relationships between the variables being examined. In anomaly detection, the Bayesian network consists of all of the variables that exist as opposed to only the variables initially contained in the process. The use of Bayesian networks for troubleshooting management is based on the initial testing data that are present, and the accuracy of those data [23].

## 4.1. Naïve Bayes

The Naïve Bayes classifier [17] operates in accordance with the Bayes rule of conditional probability. The classifier uses all of the attributes of a dataset by analysing them individually. The idea is that all of the attributes are equally important. No single attribute is more important than another attribute. At the same time, all of the attributes of the data are considered to be independent of each other [8].

## 4.2. Bayes Networks

Bayesian networks, or Bayesnet, are also known as belief probabilistic networks. It is the structure of a Bayesian network that determines the relationships and dependencies between the variables in a dataset. Because of the fact that the structure of Bayesian networks determine dependencies and relationships between variables, the use of such networks have been proposed for use in the diagnosis of faults in some cellular networks [13].

# 5. DATA COLLECTION AND DATA PROCESSING

Network faults can be classified in a variety of ways. For example, the faults can be classified based on their cause or their type. However, the most convenient network anomaly classification method is based on location. The reason for this is because network technicians and administrators need to know where faults are occurring on a regular basis so that they can determine where malfunctions are occurring, as well as the number of faults that occur in a specific location in a given period of time [24].

The work of data preparation begins with analysing and pre-processing of database features, as well as an assessment of the correctness of the data that are to be analysed. The work of data preparation involves determining the specific data that can be used to identify network malfunctions. The selection of the data to be analysed involves several considerations, such as which data are representative of the KPIs (symptoms), data availability (inputs), the attributes of the data, and establishing limits of the network anomaly detection [9][12]. The next important issue is the generation of the alarms, as well as the relationship between the KPIs and alarms. Before discussing alarms, it is appropriate to briefly explain the difference between an alarm and a fault. An alarm is an indication of a problem or issue in a network. An alarm can be thought of as the outward symptom of a problem. In contrast, a fault is the actual cause of a problem in a network.

The fault is the actual issue that needs to be corrected in order to restore a network to optimisation [2]. The alarms are generated when the nodes of the KPIs cross a specific limit or threshold that is pre-determined. It must be recognized that the thresholds are not set randomly. Instead, the thresholds are set based on their ability to provide an indication of a problem. At the same time, the threshold levels must be such that the probability of false alarms is minimised [2].

After each KPI is associated with an alarm each alarm is associated with a specific security level. An example might be a binary alarm that has the possible conditions or states or On and Off. If the KPI should be On, but it is found to be Off, then an alarm occurs. A KPI can have as many states or conditions as necessary. Some KPIs might have three states, while others might have four or even five states [2].

The KPIs are derived with the help of counters using different formulations. A single counter helps to provide a very limited indication of the larger network. However, with the use of several counters, it is possible to gain a much broader view of the network. In this study, the evaluation is presented on the basis of four major KPIs used as an input to create a classification algorithm to perform the data for the pre-processing process [13]. Table 1 shows the four KPIs that are used are Call set up success rate (CSSR), Call Drop Rate (CDR), Handover Failures (HOF) and Call Traffic Rate (TR). In addition, each KPI can take one of three states: Normal, Critical, or Warning.

Table 1. KPI metrics and alarms conditions

| State Alarm Indicator | | | |
|---|---|---|---|
| **KPI** | **Normal** (Norm) | **Critical** (Cr) | **Warning** (Warn) |
| DCR | DCR<2 | 4> DCR >=2 | DCR >=4 |
| CSSR | CSSR >=98% | 90%> CSSR =>98% | CSSR <=90% |
| TR | TR<60% | 60%> TR> =70% | TR> =70% |
| HOF | HOF<10% | 10> HOF >15 | HOF >=15 |

Table 1 also shows the specific metrics that have been established as thresholds for each alarm. The KPI metrics of Handover Failure (HOF) is the ratio of unsuccessful intersystem handovers from 3G to 2G over the total number of such attempts. The call dropped rate (CDR) is the ratio of dropped calls over the total successfully established calls [15]. The call set-up success rate (CSSR) is the rate of call attempts until successful assignment. The call drop rate (CDR) is the rate of calls not completed successfully [16]. Furthermore the traffic rate indicator is the actual rate or percentage of traffic on the network [14].

It must be noted that early symptoms of network failures are network-level errors. In this regard, several KPIs have been selected, and several conditions were identified that were deemed to most accurately reflect the problems or irregularities in the communication process control networks [21].

## 5.1. Network Fault Classification Algorithm

We selected the most popular Key Performance Indicators to create a Fault Classification Algorithm. The algorithm parameters must be set up for the input data. The rules generated for the fault classification algorithm are given bellow:

```
Input data file:
@KPIalarms,@DroppedCallsRate,@CallSuccessRate,
@TrafficRate,@HandoverFailures

do if DCR>2 AND CSSR>= 90.
-   do if DCR<=4 AND CSSR<98.
-       #KPIAlarmClass=CR.
-   else if TR> 60 AND HOF>10.
-   do if TR<70 AND HOF<25
-           #KPIAlarmClass=CR.
-   else if DCR>4 OR CSSR<90
-       do if  HOF>=25 OR TR>70.
-           #KPIAlarmClass=WARN.
-           else if KPIAlarmClass=NORM.
-           end if..

FREQUENCIES
 VARIABLES=#Data,KPI
 /FORMAT=DFREQ
 /ORDER=  Network Fault .
execute.
```

Figure 2.  Fault classification algorithm

The parameters will include equality or inequality for the state alarm indicators and greater than or less than for the KPI alarms metrics.  The relational nodes can be placed into decisions trees or into rules.  However, plans that use relations are more often placed in rules rather than in decision trees.

For test networks, normal and fault data are available.  However, for production networks, only normal data are generally available.  This creates an obstacle for network technicians and administrators because the classifier for network troubleshooting will not be well trained for the real conditions of the network environment.  This is certainly be the case if the real network is greatly different from the test network on which the classifier was trained.  In addition, without training that includes fault data for the real network, the classifier will be even more inefficient at detecting actual network faults. If these conditions are present, then additional training will need to occur so that the classifier can be effective at detecting faults on the production network [22].

Fault classification algorithms can be created to fault detection and fault localisation.  For fault detection, the algorithm is created based on all of the elements and conditions of network troubleshooting.  For fault localisation, however, the algorithm can be created to diagnosis the specific network failure that is determined to exist.  In other words, different types of data can be used for the specific needs of the network technicians and administrators.  In this way, the goal of this investigation is to develop and create a method by which different types of fault algorithms can be used.

**5.2. Explanation of KPI Metrics**

The next step in the process is the selection of a subset of instances in Weka for the parameters of KPIs shown in table 2. It is important to note that the selection of the subset of instances was not random. Instead, the selection of subsets of instances for this study was based on previous studies and investigations of fault diagnosis in networks using data mining classifiers [14][16][17].

Table 2. KPI Acronyms and abbreviations

| Acronym | Definition |
|---|---|
| BSC | Base Station Controller |
| **Radio Access Indicators** | |
| RAN | Radio Accessibility Network |
| RAB | Radio Access Bearer |
| **Traffic Channel Indicators** | |
| TCH | Traffic Channel |
| TCHCR | Traffic Channel Congestion Rate |
| TCH Availability | Mean number of available channels |
| TCHDR | Traffic Channel Drop Rate |
| TCHSSR | Traffic Channel Success Rate |
| **Standalone Dedicated Control Channel Data** | |
| SDCCHCR | Standalone Dedicated Control Channel Congestion Rate |
| SDCCHAR | Standalone Dedicated Control Channel Access Rate |
| SDCCHDropsExcessiveTA | The number of Standalone Dedicated Control Channel Drops due to Excessive Timing Advance |
| SCDDHDSuddLostCon | Standalone Dedicated Control Channel Suddenly Lost Connection |
| **Handover Indicators** | |
| HF | Handover Seizures Failures |
| HOFR | Handover Failures Rate |
| HOSR | Handover Success Rate |

The instance of BSC represents the base station controllers, while Id represents the cell Ids. The Standalone Dedicated Control Channel Congestion Rate (SDCCHCR) indicates the probability of accessing a stand-alone dedicated control channel available during call set up [14] [17]. TCH Seizure Attempts is the number of traffic channels that are allocated for traffic [17]. The SDCCH Access Rate shows the percentage of call access success rate received in the base station location [17]. TCH Availability is the rate of Traffic Channels availability in the network, while the TCH Drop is the total drop rate in the network and also the TCHCR is the rate of TCH congestion in a network area [14].

The Handover Success parameter shows the percentage of success handovers on the network of all handover attempts [15]. The SDCCH Drops Excessive TA presents the number of drops due to excessive timing advance [17]. Based on those relational nodes, the system would identify the optimised status of the TCH (traffic channel), RAN (radio accessibility network performance audit) and RAB (Radio Access Barrier) show their optimised states [17]. The Traffic Channel Congestion Rate of the SDCCH drops measures the total number of RF losses as a percentage of the total number of call attempts for the SDCCH channels on the network. The SDCCD Availability Rate is the percentage of SCDDH channels that are available on a network at a given time [14].

The SCDDHDSuddLostCon is the rate of connections that are suddenly lost on a network [17]. Additionally the TCH Drop Rate is the rate of traffic assignment failures, while TCHCR is the congestion rate related to call setup traffic [17]. The HOFR statistic is intended to give an indication of the rate as a percentage of handover failures in relation to total handovers [14]17]. The TCHSS shows the percentage of TCHs that are successfully seized [17]. The SDCCH Drops is the number of drops due to low signal strength or network congestion [14] [17]. The Handover Success Rate is the percentage of rate of handover success. Finally, the KPI Alarms are the symptoms that occur due to the establishing limits that are set by the classifiers of the system.

# 6. IMPLEMENTATION OF THE SYSTEM

The implementation of network troubleshooting may generate a large number of alarms. The problem with the generation of a large number of alarms is the potential for a high rate of false alarms. The ability for a human to visualize all of the alarms that may be generated is difficult or even impossible. In this way, the alarms generally need to be aggregated so that visualization can occur more easily for the person that is responsible for examining them [2].

Aside from the aggregation of alarms for the purpose of easier visualization, it is also possible to implement an automatic fault identification module that can correlate the observed alarms in order to indicate the root cause of the alarms. With the automatic fault identification module, a human may not even be needed as an automatic recovery system can be used to use the alarms to quickly determine the fault and correct it [2].

For this study, 2,100 instances are inserted that represent the Cell Ids and Base Station locations. In addition, 25 attributes are input that show the data and key performance indicators including the data base. The dataset representation in ARFF (Attribute-Relation File Format) is shown below:

@Period,@BSC,@Id,@SDCCHCR,@TCHSeizureAttempts,@TCHAccessRate@TCHAvailability,@TCHDrop,@TCHCR,@TCHTR,@HandoverSuccess,@SDCCHDropsExcessiveTA,@RAN,@SDCCHDR,@SDCCHAvailabilityRate,@HF,@SDCCHDropSuddLostCon,@TCHAssignmentFailureRate,@TCHDropRate,@TCHCongestionRate,@HOFR,@TCHSS,@SDCCHDrops,@HOSR@KPIAlarms

Once classification is settled down, representation of data can use data visualization techniques of decision trees and rules classifiers. Figure 3 shows the JRP optimisation rules.

The JRP optimisation rules show that six rules were extracted from Weka regarding the JPR classifier. The rules that were generated show the conditions of critical optimization faults and the KPI metrics to show when a cell can be considered to be optimized. Optimization is considered to occur when the Normal (NORM) condition is achieved. The last rule that was generated tells the system that if the cell is not classified as either normal (NORM) or critical (CR), then it needs to be checked for a warning (WARN) status.

```
JRIP rules:
===========

(TCHDropRate <= 1.96) and (TCHSeizureAttempts <= 59.78) and (SDCCHCR <= 5.57) => KPIAlarms=NORM (319.0/0.0)
(TCHSeizureAttempts <= 12.93) and (TCHDropRate <= 5) => KPIAlarms=NORM (9.0/0.0)
(TCHDropRate <= 4.99) and (RAB <= 51.09) and (TCHAvalability <= 99.22) => KPIAlarms=CR (639.0/1.0)
(TCHSeizureAttempts <= 59.14) and (RAB >= 49.52) and (TCHSS >= 90.01) => KPIAlarms=CR (126.0/1.0)
(RAB <= 15.98) and (HF >= 714) and (SDCCHCR <= 7.97) => KPIAlarms=CR (42.0/0.0)
 => KPIAlarms=WARN (965.0/1.0)

Number of Rules : 6
```

Figure 3.  JRP Optimisation rules

```
Classifier output

TCHDropRate <= 1.96 AND
RAB <= 38.91 AND
HOFR <= 6.76: NORM (314.0)

TCHDropRate > 9.98: WARN (433.0)

RAB > 51.11: WARN (230.0)

SDCCHDR > 193 AND
TCHDropRate > 3.01: WARN (103.0)

TCHCR <= 4.16 AND
TCHAvalability > 90.66 AND
TCHDropRate <= 5 AND
TCHDropRate > 2.35 AND
TCHDrop > 6.2: CR (608.0)

HOFR > 7.99 AND
TCHAvalability > 90.66 AND
TCHSS > 89.98 AND
TCHDropRate > 1.6: CR (179.0/1.0)

TCHDropRate > 5: WARN (190.0)

TCHCR <= 3.46 AND
TCHAvalability <= 98.99: CR (20.0)

RAB <= 30.32: NORM (14.0)

: WARN (9.0)
```

Figure 4. Part optimisation decision list

Figure 4 shows the optimisation decision list.  A total of six rules were extracted from Weka regarding the Part classifier.  As before, the rules that were generated illustrate the conditions of critical optimization faults and the KPI metrics to show when a cell can be considered to be optimized.  Optimization is considered to occur when the Normal (NORM) condition is achieved. The last rule that was generated tells the system that if the cell is not classified as either normal (NORM) or critical (CR), then it needs to be checked for a warning (WARN) status.  The rule based detection system now expects an alarm to occur after other alarms have been triggered. The rules that indicate the normal (NORM) classification shows the records that need to be changed or adjusted so that the network can be returned to an optimised state.

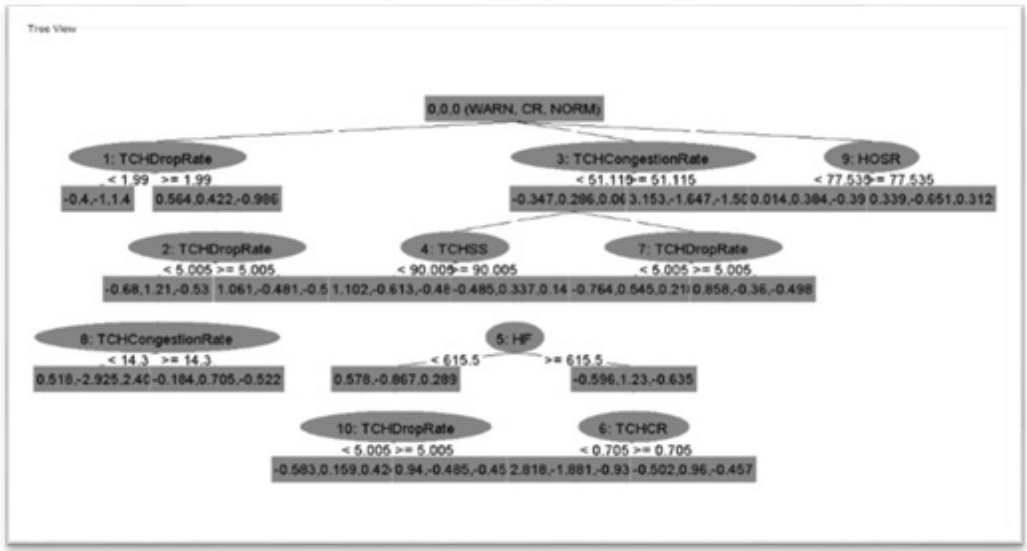

Figure 5.  Optimisation decision LADtree

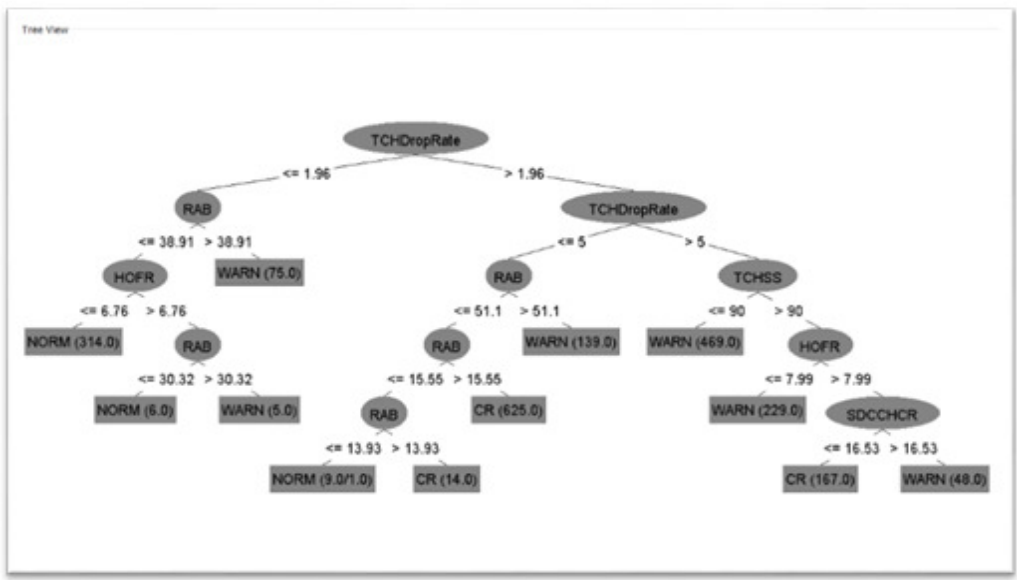

Figure 6.  Optimisation decision tree J48

Figures 5 and 6 show the optimization decision trees for J48 and LAD Tree.  The number of Leaves is 12 for J48 class and the Size of the tree is 23.  The attribute selection from the J48 and LADTree classifiers is based on the variables related to the utilization of the network such as Traffic Channels and Handover success and Number.  The LAD Tree provides a Logical Analysis of Data as the classification method.  The LAD Tree shows a prediction for anomaly detection that is based on 21 predictor nodes at the tree side and 14 predictor nodes for the optimisation results at the leaves of LAD Tree [11].

The Weka trees discovered rules and results of key performance indicators alarm predictions based on those rules respectively.  Perhaps just as importantly, the decision trees demonstrate the ability to provide a visual representation of the faults that can be understood by a human operator whose job it might be to analyse the data and determine the cause of the lack of network optimisation.  By examining the decision trees, it is possible to quickly determine which cells are indicated as being normal (NORM) and which cells are indicated as being critical (CR).  By looking at the specific classifiers, the human operator can truly make a quick decision about the fault in the network to correct that fault and return the network to maximum optimisation.

As has already been noted, network anomaly detection can be classified in different ways, but the most convenient and useful classification for network technicians and administrators is by location.  By grouping the faults based on location it is possible for the technicians and administrators to not only know where faults are occurring regularly, but how many issues are occurring in specific locations at given points in time and over specific time intervals [24].

Furthermore, group faults that occur due to increases in reporting by users is of interest in this investigation.  In this regard, the effort has been made to match all available data that were explicitly collected or that were received based on sensed signals in a way that created a logical cause and effect relationship.  Then, those data and relationships were compared to the logical network.  The output of these calculated comparisons is output probabilities.  The calculations were conducted using Bayesian networks.

Naive Bayes Classifier

| Attribute | Class WARN (0.46) | CR (0.38) | NORM (0.16) |
|---|---|---|---|
| Period | | | |
| 16/06/2014 | 7.0 | 147.0 | 2.0 |
| 20/06/2014 | 41.0 | 174.0 | 1.0 |
| 17/06/2014 | 3.0 | 1.0 | 1.0 |
| 15/06/2014 | 2.0 | 1.0 | 1.0 |
| 18/06/2014 | 3.0 | 1.0 | 1.0 |
| 14/06/2014 | 2.0 | 1.0 | 1.0 |
| 19/06/2014 | 2.0 | 1.0 | 1.0 |
| 13/06/2014 | 80.0 | 40.0 | 50.0 |
| 21/06/2014 | 40.0 | 93.0 | 1.0 |
| 24/06/2014 | 2.0 | 2.0 | 1.0 |
| 25/06/2014 | 2.0 | 2.0 | 1.0 |
| 26/06/2014 | 2.0 | 2.0 | 1.0 |

| Attribute | WARN | CR | NORM |
|---|---|---|---|
| BSC | | | |
| BSC2 | 3.0 | 2.0 | 4.0 |
| BSC5 | 25.0 | 2.0 | 1.0 |
| BSC411 | 2.0 | 1.0 | 1.0 |
| BSC1 | 1.0 | 2.0 | 2.0 |
| BSC21 | 80.0 | 47.0 | 3.0 |
| BSC25 | 7.0 | 6.0 | 2.0 |
| BSC55 | 6.0 | 6.0 | 2.0 |
| BSC18 | 3.0 | 4.0 | 1.0 |
| BSC05 | 1.0 | 1.0 | 2.0 |
| BSC11 | 54.0 | 96.0 | 56.0 |
| BSC19 | 3.0 | 4.0 | 1.0 |
| BSC10 | 8.0 | 5.0 | 3.0 |
| BSC293 | 2.0 | 1.0 | 1.0 |

Figure 7.  Fault localization

Figure 7 shows the fault localization by time and base station controller using Bayesian classifiers. The Naïve Bayes classifier shows the fault localisation and probability distributions of the alarms by base station location and day. The classifier uses all of the attributes of the dataset by analysing them individually. This was done to show how the information the operators would use to find the cause of the faults. It is important to explain that three nodes were used with the Bayesian classifiers, which were sensor status, process variable, and measured value. It is also possible to achieve fault detection based on the probabilities for the sensor status node states. However, one-to-one mapping between the nodes is not always possible. In order to carry out simultaneous detection of both sensor and process faults, additional nodes have to be used in the model, which requires knowledge of the faults and their impacts on the process variables. Unfortunately, that information may not always be available [23].

# 7. PERFORMANCE EVALUATION WITH SIMULATION DATA

While the goal of this investigation was to create a method of automated troubleshooting management using data from multiple networks, the failure detection capabilities of the method with traffic data only was also examined. In order to examine failure detection capabilities with traffic data, three specific network faults were difficult to detect were simulated. Then, the detection method was applied to the traffic data [26].

## 7.1 Re-Classification of Network Faults

With the three traffic patterns of network faults that were determined to be difficult to detect, an evaluation occurred as to whether the anomaly-detection method could actually detect those patterns and faults. What is needed is to re-classify the network faults based on the normal (NORM) optimisation rules in order to correct the network faults [26].

# 8. FAULT DIAGNOSIS RESULTS DISCUSSION

An average of 860 cells of the data base were classified in a state alarm of "Warning" , 785 found to be "Critical" and only 320 cells were considered as "Normal". The most popular KPI metrics extracted by the system are based on TCH Availability, Congestion and Success Rate, Handover Success and Failure Rate, SDCCH and TCH Drop Rate and final TCH attempts.

## 8.1. Evaluation Metric Using Different Classifiers

In order to measure the accuracy of the decision trees that were generated for the network fault classification results, three measures were used that are widely used to evaluate data mining techniques, which are precision, recall, and F1 metric. The precision and recall measures are calculated based on true positive (TP) or the number of objects that are correctly classified, false positive (FP) or the number of classifiers that are not correctly classified, and false negative (FN), or the number of positive objects that are incorrectly classified as other classes. Precision is calculated as the ratio of true positive (TP) to the sum of true positive (TP) and false positive (FP). Recall is the measure of how many objects in a class are misclassified, and is the ratio of true positive (TP) to the sum of true positive and false negative (FN). The F1 metric is the harmonic mean of recall and precision [21].

Table 3.  Performance measure results

| Weka classifiers | Correctly classified instances | Incorrectly classified instances | Kappa statistics |
|---|---|---|---|
| J48 | 2099 | 1 | 0.99 |
| LADTree | 2097 | 3 | 0.99 |
| JRIP | 2097 | 3 | 0.99 |
| Part | 2099 | 1 | 0.99 |
| BayesNet | 1839 | 261 | 0.80 |
| NaiveBayes | 1701 | 399 | 0.69 |
| **Weka classifiers** | **Mean absolute error** | **Root mean squared error** | **Accuracy** |
| J48 | 0.0006 | 0.01 | 99.9% |
| LADTree | 0.0243 | 0.05 | 99.8% |
| JRIP | 0.0019 | 0.03 | 99.8% |
| Part | 0.0006 | 0.01 | 99.9% |
| BayesNet | 0.08 | 0.28 | 87.5% |
| NaiveBayes | 0.13 | 0.34 | 81% |

Table 3 also shows the accuracy alarm prediction by the proposed technique.  The table shows that the performance measure results of J48 and PART classifiers have the same performance in terms of accuracy in classification.  Both methods achieved an accuracy rate of 99.9%.  With the rules and Decision trees classifiers, the accuracy rate is actually more than 99.9%.  However, the Bayesian classifiers showed a decline in their accuracy levels as compared to the other methods [12].

This scale by which to measure the performance of algorithms is important in relation to the method that was tested in this study because distributed fault detection and isolation with the use of a Bayes classifier has demonstrated that even if the accuracy is not too high, it is the total output of the use of the classifier that provides important information that can be used by wireless providers to isolate and localize the network malfunctions.

Table 4 shows the evaluation of final statistics.  The scale is based on the level of accuracy of the algorithm with regards to correctly identifying faults as compared to false results [12].  The final statistics show the accuracy of the model and visualize precision and recall for ROC curve analysis (true positive rate vs false positive rate). [10].  TP rate also demonstrates the sensitivity of the model such as a scale of actual positive values and the FP rate shows the specificity such as the negative tuples that are incorrectly labelled. [11].  In this regard, the slightly less accurate information of BayesNet and NaiveBayes classifiers across the entire system still results in a high level of ability to determine and localise the fault or faults that need to be corrected in order to return a network to optimisation [12].

Table 4.  Comparison of final statistics weighted average

| Weka classifiers | TP Rate | FP Rate | Precision | Recall |
|---|---|---|---|---|
| J48 | 0.99 | 0.00 | 0.99 | 0.99 |
| LADTree | 0.99 | 0.00 | 0.99 | 0.99 |
| JRIP | 0.99 | 0.00 | 0.99 | 0.99 |
| Part decision list | 1.00 | 0.00 | 1.00 | 1.00 |
| BayesNet | 0.87 | 0.03 | 0.91 | 0.87 |
| NaiveBayes | 0.81 | 0.10 | 0.84 | 0.81 |
| Weka classifiers | F-Measure | MCC | ROC Area | PRC Area |
| J48 | 0.99 | 0.99 | 0.99 | 0.99 |
| LADTree | 0.99 | 0.99 | 0.99 | 0.99 |
| JRIP | 0.99 | 0.99 | 0.99 | 0.99 |
| Part decision list | 1.00 | 0.99 | 1.00 | 1.00 |
| BayesNet | 0.88 | 0.83 | 0.99 | 0.98 |
| NaiveBayes | 0.80 | 0.70 | 0.93 | 0.90 |

## 9. CONCLUSIONS

This paper proposes a systematic approach for anomaly detection that is based on a KPI data analysis model using the Weka environment for those who are responsible for the monitoring and optimisation of GSM networks.  The importance of the method of anomaly diagnosis that has been proposed in this paper is that it is one that has not been widely studied.  In this regard, a potentially new method, or at least one that has not been greatly considered, is available for network operators who want an automated troubleshooting system by which to quickly identify faults and return networks to optimisation in order to provide a high quality of service to users.

The state alarm indicator values shown in the method used in this study provide a strong means by which to perform network diagnosis efficiently and quickly.  The rules and decision trees set up that are extracted by the data mining process simplify the optimisation process.  The decision trees allow network operates to visually examine multiple network parameters, such as traffic channels and handovers statistics.  The end result is that the task performance management network optimisation through fault location finding, and root cause analysis is much easier [6].

A weakness of the system as this point is that the decision trees do not always show the causes of the symptoms that are creating inefficiency in a network.  However, even with this weakness, the decision trees provide the rules of the alarms, and have a high rate of accuracy in the information that is provided. In this regard, this method can be used for identifying the process fault as a means of learning detection rules from the KPI metrics.

The temporal tree was loaded so that the rules could be generated and evaluated with the decision tree generator. The rules that were found were to find the specific alarm that was to be triggered when an initial alarm occurred, and the time interval for which it was to occur. In this way, the temporal tree allowed for information about consecutive alarms that would occur after an initial alarm. In this way, the temporal tree did work effectively for symbol prediction and rule generation. Even more, the temporal tree was effective with both the training data and the actual test data [27].

It is important to iterate the method of network troubleshooting shown in this paper is not only highly accurate, but it is also efficient. A method of fault diagnosis can be high, but a lack of efficiency in using the method can be ineffective for network operators who must return a network to optimum performance levels in a matter of seconds or minutes, rather than hours [18]. Wireless networks have become more complex with increasing numbers of users [19]. This only adds to the burden of finding automatic methods of fault diagnosis that are efficient, both in terms of accuracy of output and actual usage.

The method of fault diagnosis described in this paper is indeed computationally efficient regarding the rules, decision trees and Bayes classifiers. It also represents optimisation solutions defined by rules. And facilitate the correction of optimisation faults techniques.

Furthermore, this method of fault diagnosis retains the ability to identify individual KPI metrics and alarm factors severely affecting reliability of fault location calculation. In the end, the results presented in this paper allow for the conclusion that the proposed method has produced very promising results in the classification of multi-class optimisation faults.

## 10. FUTURE WORK

The results of this study are not only important in terms of what can be concluded from them about the method of network fault diagnosis that was presented, but also the broader context of network fault management. The analysis of network fault diagnosis in this study, and most other studies, requires some knowledge of statistical analysis. The question that arises, however, is whether the people who are responsible for fault diagnosis in practice have the statistical analysis that is needed to efficiently perform the same procedures and analysis conducted in this study and others. While the focus of so much research in this area has been on methods of fault detection, it might be time to study those who have to implement these methods and use them to make real-world decisions.

Another area for future research might be to examine the business decisions that underlie network optimisation efforts on the part of network providers. While the focus of so much of this area of research is on the techniques and methods, the reality is that there are important business decisions that dictate which techniques and methods are used, as well as how they are used [20]. It would be useful to understand the decision-making process and the business realities that determine the specific methods that companies that operate wireless networks use to perform fault diagnosis.